# Gauge invariance in phase-gradient metagratings

Qingjia zhou, Lei Gao[*] and Yadong Xu[†]

[1]*School of Physical Science and Technology, Soochow University, Suzhou 215006, China.*

[2]*Jiangsu Key Laboratory of Thin Films, Soochow University, Suzhou 215006, China.*

**Abstract:** Phase-gradient metagratings/metasurfaces (PGMs) have provided a new paradigm for light manipulations. In this work, we will show the existence of gauge invariance in PGMs, i.e., the diffraction law of PGMs is independent of the choice of initial value of abrupt phase shift that induces the phase gradient. This gauge invariance ensures the well-studied ordinary metallic grating that can be regarded as a PGM, with its diffraction properties that can fully predicted by generalized diffraction law with phase gradient. The generalized diffraction law presents a new insight for the famous effect of Wood's Anomalies and Rayleigh's conjecture.

**Introduction**

Recently phase-gradient metagratings or metasurfaces (PGMs) have been reported as a new way for the manipulation of light or electromagnetic (EM) wave propagation, leading to a large number of new effect or physics [1-4]. These include efficient focusing [5], ultrathin cloaking [6], photonic spin Hall effect [7], metalens [8], wavefront control [9-11], and others [12-16]. PGMs are periodic arrays of a carefully designed supercell with *m* unit cells (*m* is integer number) that discretely introduce a covering $2\pi$ abrupt phase shift (APS) $\phi(r)$ along the surface, yielding a phase gradient $\nabla\phi(r)$ that is physically equivalent to the wave vector [1]. This phase gradient modifies the Snell's law, one of the fundamental laws of optics, leading to a generalized reflection/refraction law [1]. Essentially, the fundamental physics behind the PGMs is phase gradient induced by the APS. Here, we will show for the first time that the phase gradient is independent of the initial choice value of APS. So, for a fixed phase gradient, the initial value of APS in PGMs is not unique and can be arbitrary, which predicts the existence of gauge invariance in PGMs. Such gauge invariance provides a new insight to revisit the ordinary metallic grating (MG) [17] that is well studied in nanophotonics and plasmonics.

    Wood's anomalies are well-known effect in nanophotonics, which were first discovered by Wood in 1902 in experiments on reflection-type MG [18], and have been investigated and attracted lots of attention of scientists for more than a century [19-22]. They have obvious sudden and intense variations in the reflectance/transmittance of various diffracted orders in certain narrow frequency bands or alternatively in a certain narrow range of incident angles for a fixed operating frequency [19]. They are termed as an anomaly as the ordinary grating theory cannot explain them well. Various efforts [20-22] have been performed over the years to understand Wood's anomalies. For instance, Ugo Fano contributed them to the excitations of surface plasmons in periodically corrugated metal interface [21]. Rayleigh proposed a well-known interpretation, based on his conjecture that these anomalies occur at the wavelengths at which a diffracted order appears or disappears at a grazing angle [22]. Specifically, for the *n*[th] diffracted order, it is given by,

$$\lambda_n = (\pm 1 - \sin\theta_{\mathrm{in}})p/n, \quad (1)$$

where *p* is the period of grating, $\theta_{\mathrm{in}}$ is the angle of incidence, and *n* is an integer. The Rayleigh conjecture was considered as a valuable tool for the prediction of Wood's anomalies, thereby these anomalies are also termed as Wood-Rayleigh (WR) anomalies. These anomalies are successful in many cases, but the physical reason of the conjecture is still unclear. In particular, there is a lack of reasonable explanation of why the passing-off of the order *n* should occur, i.e., $\sin\theta_n = \pm 1$ in Eq. (1).

---

[*] leigao@suda.edu.cn

[†] ydxu@sdua.edu.cn

In this work, we start from the gauge invariance of PGMs, and then revisit the widely studied MG from the concept of PGMs and present a new insight for the Rayleigh's conjecture. We will show that the ordinary grating theory cannot accurately describe MG's diffractions. By contrast, only the new diffraction law including the phase gradient, can be used to fully figure out their diffraction features. In the new diffraction law, the diffraction order of *n*=1 is the lowest order, while the *n*=0 order is a higher diffraction order and is difficult to be coupled. This result completely contrasts the ordinary understanding of diffraction law and fully explains the conjecture of WR anomalies as shown in Eq. (1). More importantly, we also show that although the WR anomalies can be seen in any MGs, it is the most obvious when the phase gradient condition is satisfied in the MG. Our findings provide a new way to study the physics in metallic gratings from the concept of PGMs, bridging the gap of two fields of optical metasurfaces and plasmonics.

**Model and Theory**

A typical reflection-type PGM with *m*=2 is shown in Fig. 1(a), a textured metal grating made of a periodic repeated supercell with a period of *p*. Each supercell contains two unit cells of different grooves. The width of each unit cell is $\Lambda = p/2$, the width of both grooves is *w*, and their depth is $h_1$ and $h_2$, respectively. The whole structure is infinite along the *y*-direction. In order to discuss the essential mechanism, the metal is assumed to be a perfect electric conductor (PEC). Similar results can be found if the PEC is replaced with real metals. Consider a TM (i.e., only magnetic field along *y*-direction) polarized light obliquely incident from the air onto this PGM, with an incident angle of $\theta_{in}$. When light enters into the *j*th (*j*=1, 2) groove and travels along with it, as shown in Fig. 1(b), the reflected wave will get phase retardation $\varphi_j = 2\beta h_j$ at the interface (*y*=0), where $\beta$ is the wave vector of an EM wave in the grooves. As mentioned above, the concept of PGM requires a full $2\pi$ phase shift in a supercell, which shows a phase difference $\Delta\varphi = 2\pi/m$ between two adjacent unit cells. In this case, *m*=2 leads to $\Delta\varphi = \varphi_2 - \varphi_1 = \pi$, which can be achieved by adjusting the groove depth [23, 24]. Physically, the diffraction properties of the designed PGM are governed by [25]:

$$k_0 \sin\theta_r = k_0 \sin\theta_{in} + \kappa + vG, \quad (2)$$

where $G = 2\pi/p$ is the reciprocal vector and $\kappa = \Delta\varphi/\Delta x = 2\pi/p$ is the phase gradient. They share the same value, but with different physical meaning [26]. $\theta_r$ is the angle of reflection. Moreover, the diffraction of PGM can be re-expressed as,

$$k_0 \sin\theta_{r,n} = k_0 \sin\theta_{in} + nG, \quad (3)$$

where $n = 1 + v$ is the diffraction order. Note that Eq. (2) is very similar in formula to ordinary grating diffraction equation, but is quite different in physics from it because the phase gradient makes the lowest order corresponding to *v*=0.

In fact, the initial value of APS is not unique for the design of PGMs. If we take the following transformation,

$$\phi'(r) \to \phi(r) + \Phi_0, \quad (4)$$

where $\Phi_0$ is a constant, then $\nabla\phi'(r) = \nabla\phi(r)$. It implies that when the APS of all spatial positions increases by the same value, the behaviors of PGMs and the associated diffraction law of Eq. (2) are completely unchanged. The reason behind this is the constant phase gradient of

PGMs. This is a global gauge invariance of PGMs, which is similar to that of the electric potential of an electrostatic field, requiring the choice of zero point.

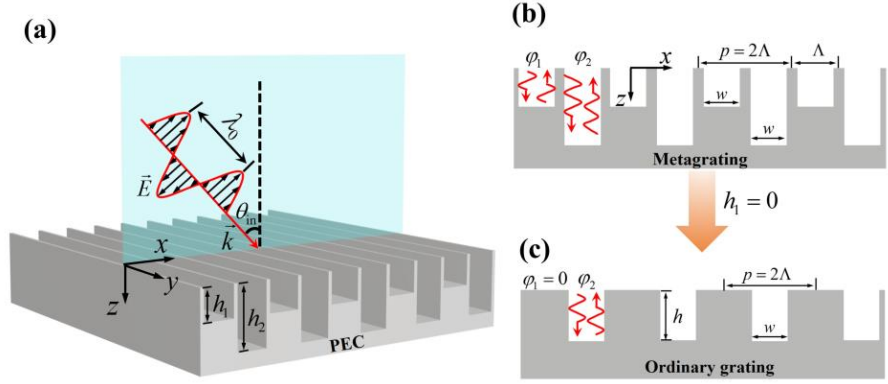

**Fig. 1.** (a) Schematic diagram of the reflection-type metagrating in air, where one period is composed of two air grooves with depth $h_1$ and $h_2$. A TM light incident to metagrating with an incident angle $\theta_{in}$ and wavelength $\lambda_0$. (b) *x-z* plane of (a). A supercell with period *p* contains two identical unit cells with width $\Lambda$. The width of groove in each unit cell is *w*. (c) For $h_1 = 0$ in (b), the metagrating becomes an ordinary grating.

With this global gauge invariance in PGMs, we choose $\varphi_1 = 0$ in PGM as shown in Fig. 1(b), where $h_1 = 0$. Then the PGM becomes an MG, as seen from Fig. 1(c), and both cases should share the same diffraction law of Eq. (2). In this way, the concept of PGM can give more profound physics to the diffraction properties of MG than that from ordinary diffraction theory. As we know, ordinary diffraction theory describes the lowest diffraction order of *n*=0, and it is preferred to be coupled and diffracted in the process of grating diffraction. From the view of PGMs, however, the lowest diffraction order of the MG is not *n*=0, but *n*=1 corresponding to $v = 0$ in Eq. (2), which is preferred to be diffracted. More importantly, Eq. (2) tells us that *n*=0 is a higher diffraction order and more difficult to be coupled than that of order *n*=1 [25].

In order to further reveal the diffraction feature, theoretical analyses were performed for the designed PGM. The total magnetic field in air region ($z \geq 0$) can be expressed as the sum of the incident and reflected fields, $H_{y,1} = \sum_n \left[ \delta_{n,0} \exp(ik_{z,n}z) + r_n \exp(-ik_{z,n}z) \right] \exp(i\alpha_n x)$, where $\delta_{n,0}$ is the Kronecker delta function, $k_{z,n} = \sqrt{k_0^2 - \alpha_n^2}$ is the z-component of wave-vector of the $n^{th}$ order diffracted wave, $\alpha_n = k_0 \sin\theta_{in} + nG$ is the wave vector along the *x*-direction, and $r_n$ is reflection coefficient of the $n^{th}$ diffraction order. As the subwavelength grooves, we simply assume that only fundamental mode exists inside the grooves with propagating wave vector $\beta = k_0$. In this way, the inside magnetic field of the $j^{th}$ groove ($-h_j \leq z < 0$) is given by $H_{y,2j} = a_j \exp(ik_0 z) + b_j \exp(-ik_0 z)$, where $a_j$ and $b_j$ are the amplitude coefficients of the forward and backward waves. In addition, the corresponding electric fields of each region can be analytically obtained with the help of $\nabla \times \mathbf{E} = -\partial \mathbf{B}/\partial t$. By applying the continuous boundary conditions of electric and magnetic fields at $z = 0$ and $z = -h_j$, we obtain the following equations,

$$\sum_n g_n(\delta_{n,0} + r_n) \exp(i\alpha_n x_j) = b_j \left[ 1 + \exp(2ik_0 h_j) \right], \quad (5)$$

$$\sqrt{1 - (\alpha_n/k_0)^2} \left[ \delta_{n,0} - r_n \right] = \sum_j (a_j - b_j) f g_n \exp(-i\alpha_n x_j), \quad (6)$$

where $x_j$ is coordinates of the $j^{\text{th}}$ groove, $f = w/p$ is the filling factor, and $g_n = \operatorname{sinc}(\alpha_n w/2)$. By solving Eq. (5) and Eq. (6), we can obtain all order reflection coefficients $r_n$, where reflectivity is $R_n = |r_n|^2$.

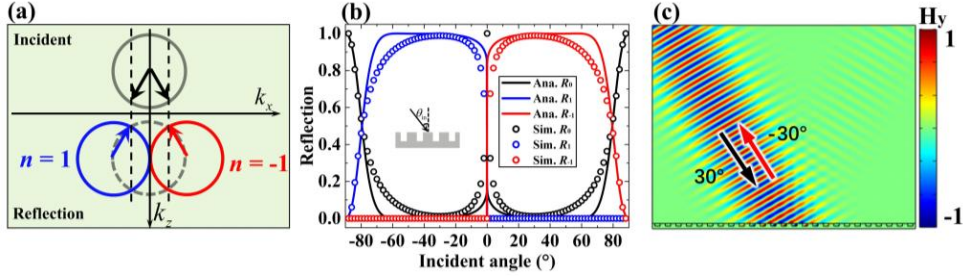

**Fig. 2** The case of $\kappa = k_0$. (a) The iso-frequency contours for wave diffraction in PGM. The gray circle is for the incident wave with the black arrows indicating the incident direction. The blue circle, gray dash circle, and red circle represent reflected wave of $n=1, 0, -1$ diffraction orders, respectively. The direction of $n=1$ and -1 order reflected waves are denoted by blue and red arrows, respectively. (b) The relationship between reflectivity and incident angle of all diffraction orders. Solid lines and hollow circles represent analytical and simulated results, respectively. (c) Magnetic field distribution for a nearly perfect retroreflection. The incident angle is 30° and reflection angle is -30°.

To clearly illustrate our idea, we first consider a PGM with $p = \lambda_0$ that corresponds to $\kappa = k_0$. The geometric parameters are $h_1 = 0$, $h = h_2 = \lambda_0/4$ and $w = p/2$. The designed PGM is a MG with a filling ratio of $f = 0.5$, where WR anomalies can be observed in the spectra (see Supplemental document). Because they are equivalent, so, the diffractions in MG can be governed by Eq. (2), i.e.,

$$\sin\theta_{r,n} = \sin\theta_{\text{in}} + 1 + v. \tag{7}$$

In Eq. (7), the order $v = 0$ is the lowest diffraction order that corresponds to $n=1$ and defines a critical incident angle of $\theta_{\text{in}} = 0$. For $\theta_i < 0$, then $\sin\theta_{r,1} = \sin\theta_{\text{in}} + 1$. While $\theta_{\text{in}} > 0$, higher-order diffraction with $v \neq 0$ happens. The iso-frequency contours of both incidences and the reflection of all possible diffraction channels are shown in Fig. 2(a). It is found that when $\theta_{\text{in}} > 0$, the higher-order such as $v = -1$ and -2 are obtained that correspond to $n=0$ and $n=-1$, respectively. Considering the case of these two high diffraction orders, the ordinary diffraction theory describes that $v = -1$ ($n=0$) can be preferably coupled than that of $v = -2$ ($n=-1$). However, this is not possible due to phase gradient. In fact, the higher order of diffraction follows a new rule of $L=m+1-n$ [25], which indicates that the higher the diffraction order, the higher will be the coupling and diffraction. This statement strongly contrasts the ordinary diffraction theory which reveals that the lower the diffraction order, the higher will be the coupling priority. Therefore, in current case the $n=-1$ order is preferred to be coupled compared to $n=0$ order. These results are further validated and confirmed by the diffraction efficiency of each orders. Figure 2(b) displays the relationship between the reflectivity of the $n=0, 1$, and -1 diffraction orders and the incident angle. In this plot, solid lines represent the analytical results while the circles depict the simulated results, obtained from COMSOL Multiphysics. The analytical results agree well with the simulated results each other. In particular, when $\theta_{\text{in}} = 30°$, the reflection is mainly dominated by $n=-1$ order rather than $n=0$ order. The field pattern of an incidence $\theta_{\text{in}} = 30°$ is shown in Fig. 2(c) which clearly describes the retroreflection [11] at $\theta_r = -30°$ corresponding to $n=-1$.

Similar results are obtained for other phase gradients, i.e., $\kappa = 1.5k_0$. Figure 3(a) shows the corresponding iso-frequency contours, where two lowest diffraction order $n=1$ (blue circle) and $n=-1$ (red circle) are separated. In this case, the period is $p = 2\lambda_0/3$, which indicates the abrupt change in diffraction efficiency (i.e. WR anomaly) happening at $\theta_{in} = 30°$, predicted by Eq. (1). This transition angle can be clearly seen from the calculated diffraction efficiency of each order, as shown in Fig. 3(b). As the incident angle increases from zero and when it crosses $\theta_{in} = 30°$, the dominant diffraction order instantaneously changes from $n=0$ order to $n=1/-1$ order, although the $n=0$ order channel is still open. A nearly perfect retroreflection happens when $\theta_{in} = \pm 48.6°$ (see Fig. 3(c)), where the efficiency of $n=1$ and 0 order is 90 and 10%, respectively. Moreover, when $\kappa > 2k_0$ ($p < \lambda_0/2$), only the $0^{th}$ diffraction order channel is open for the incident wave, with the $n=1$ and $-1$ orders that go out of the range of $(-k_0, k_0)$. In this case, the PGM is just like a perfect mirror, with $R_0 = 1$ for all incidences (see Supplemental document). So, no WR anomalies are observed. These results are also consistent with the prediction of Eq. (1).

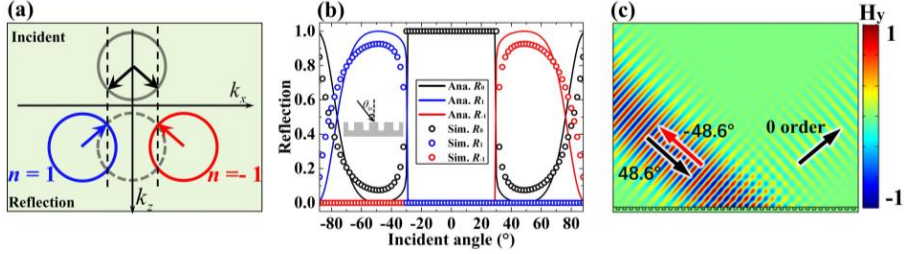

**Fig. 3.** The case of $\kappa = 1.5k_0$. (a) The iso-frequency contours for wave diffraction in PGM. (b) The relationship between reflectivity of $n=-1$, 0, 1 orders and the incident angle. (c) Magnetic field distribution for nearly perfect retroreflection. The incident angle is 48.6°. The $n=-1$ order is dominant in the reflected wave.

It is believed that WR anomalies can be predicted for any groove depth, but their intensities are closely related to the depth of the grooves. Here we find that when the APS along the interface covers full $2\pi$, i.e., it satisfies the condition of PGMs, the intensity of WR anomalies becomes strongest. In order to understand this point, Fig. 4 shows the diffraction efficiency (i.e. reflection) of $n=-1$ (Fig. 4(a)) and $n=0$ (Fig. 4(b)) order as a function of the incident angle and the groove depth for $\kappa = 1.5k_0$. Here, only $R_{-1}$ and $R_0$ are shown as diffraction efficiency of other orders are zero for $h \in [0, \lambda_0]$ and $\theta_{in} \in [0, 90°]$. It is found that when $h = 0.25\lambda_0$ and $0.75\lambda_0$ (the dashed lines in Fig. 4(a)), where the APS condition is satisfied, the diffraction efficiency of $n=-1$ is strongest. For other depth that deviates from these ideal values, the $\Delta\varphi$ differs greatly from the ideal value of $\pi$, leading imperfect APS along the interface. But due to the tolerance in designing PGM [27], the Eq. (2) still holds when the depth deviation is not very large. On the contrary, the $n=0$ order efficiency is constant for all incident angle when $h = 0.5\lambda_0$ (see the dashed line in Fig. 4(b)). No WR anomalies can be observed even if the condition of Eq. (1) is satisfied. Therefore, we conclude that the WR anomalies are related to the phase gradient.

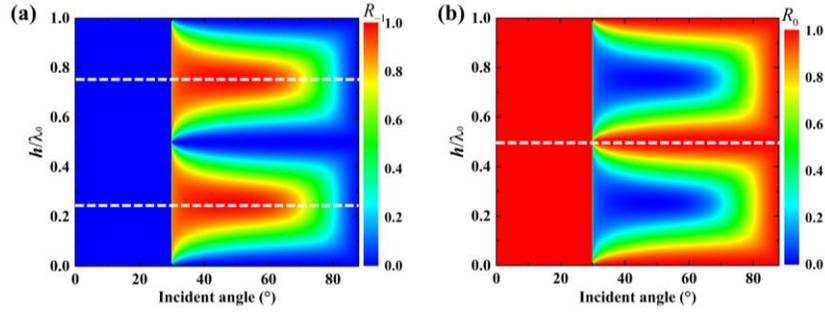

**Fig. 4.** Analytical reflectivity as a function of $\theta_{in}$ and $h$ for $\kappa = 1.5k_0$. (a) Reflectivity of $n$=-1 order. When $h = 0.25\lambda_0$ and $h = 0.75\lambda_0$ (the dash lines), the diffraction of $n$=-1 order is dominant for $\theta_{in} \in [30°, 70°]$. The abrupt change of diffraction efficiency at $\theta_{in} = 30°$ corresponds to WR anomalies. (b) Reflectivity of $n$=0 order. When $h = 0.5\lambda_0$ (the dash line), the diffraction efficiency of $n$=0 order is almost unity.

## Conclusions

In summary, we have revealed the existence of gauge invariance in phase-gradient metagratings. This gauge invariance can convert the well-studied metallic grating (MG) to PGM with $m$=2. So, the diffraction effect of MGs can be fully predicted from the generalized diffraction law with phase gradient. In particular, we have shown that the generalized diffraction law provides new insight to Wood-Rayleigh (WR) anomalies, and the physics of Rayleigh conjecture is due to phase gradient. So, the revealed gauge invariance and our results build a bridge between the fields of metasurfaces and plasmonics.

## Acknowledgments

This work was supported by The National Natural Science Foundation of China (Grant Nos. 11974010, 11774252 and 92050104), and the project funded by the China Postdoctoral Science Foundation (Grant Nos. 2018T110540), and the Priority Academic Program Development (PAPD) of Jiangsu Higher Education Institutions.

# *Supplemental document*

## 1. WR anomalies in ordinary metallic grating

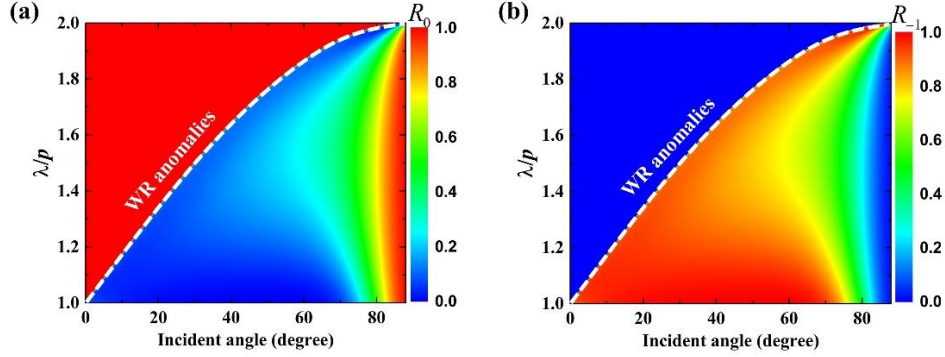

Fig. S1. WR anomalies happening in ordinary metallic grating shown by Fig. 1(c) in the main text. (a) and (b) are the calculated reflectivity of n=0 order and n=-1 order, as a function of the incident angle $\theta_{in}$ and the wavelength normalized by the period $p$. In analysis, $p$ is a fixed value, $w=p/2$, and $h=p/4$. In both plots, the white dash lines shows the positions where the WR anomalies occur.

For a 2D PEC grating as shown in Fig. 1(c), diffraction efficiency of each order can be obtain by solving Eqs. (5) and (6). Fig. S1(a) and (b) show the reflectivity of n=0 order and -1 order, as a function of the incident angle $\theta_i$ and the working wavelength $\lambda$. The period $p$ of grating is a fixed value, $w=p/2$, and $h=p/4$. Only $R_0$ and $R_{-1}$ are shown here because the diffraction efficiency of other orders are zero for $\lambda \in [p, 2p]$ and $\theta_{in} \in [0, 90°]$. WR anomalies are marked with white dash lines and well predicted by Eq. (1).

## 2. The PGM (MG) with a large gradient, behaving as a perfect mirror.

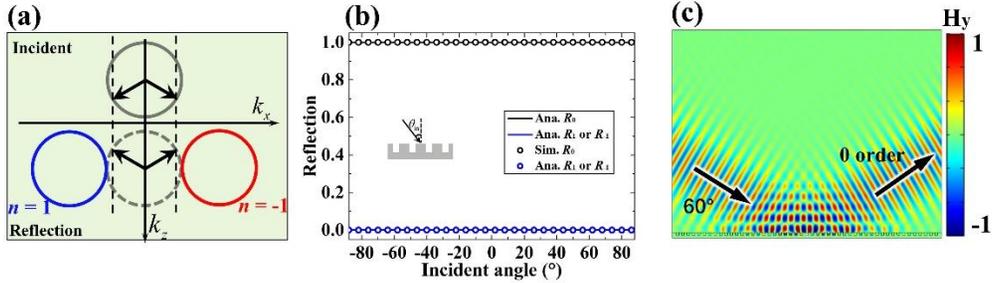

Fig. S2. The performance of PGM (MG) with a phase gradient of $\kappa = 2k_0$. (a) Iso-frequency contours for wave diffraction in the PGM. The red and blue circles representing the $n=-1$ and $n=1$ diffraction order, respectively, are separated and outside the range of $(-k_0, k_0)$. (b) Reflectivity of 0 and $\pm 1$ orders vs incident angle. Because of $R_1 = R_{-1} = 0$ for all incident angle, we only use blue solid lines and blue circles to represent the theoretical and simulated results of $R_1$ or $R_{-1}$. (c) Out of plane magnetic field distribution for $\theta_{in} = 60°$. Only 0 order exist with reflection angle 60°.

Because the phase gradient is $\kappa = 2k_0$, the 1 order circle (blue) and -1 order circle move out the range of $(-k_0, k_0)$, as shown in Fig. S2(b). Fig. S2(b) illustrates the efficiency of $n$=0 and

$n = \pm 1$ diffraction efficiency change as incident angle. Only 0 diffraction order channel is open for the incident wave. The grating just like a perfect metal plane and $R_0 = 1$ for all incident angle. The $\pm 1$ diffraction channels are closed owning to wave vector mismatch for any incident angle. Take $\theta_{in} = 60°$ as an example, Fig. S2(c) shows the Gaussian wave simulation of field pattern. The incident wave is specular reflected and the grating just likes a PEC plane.